\documentclass[aps,prl,preprint,nopacs,superscriptaddress, showpacs]{revtex4}
\usepackage{chemformula} 
\usepackage[T1]{fontenc} 
\usepackage{graphicx}
\usepackage{verbatim}
\usepackage{mathrsfs}
\usepackage{gensymb}
\usepackage{amsmath,amssymb,amsfonts,bm}

\begin{document}

\title{Observation of Symmetry-Protected Dirac States in Nonsymmorphic $\alpha$-Antimonene}

\author{Qiangsheng~Lu}
\affiliation {Department of Physics and Astronomy, University of Missouri, Columbia, Missouri 65211, USA}
\author{Kyle~Y.~Chen}
\affiliation {Rock Bridge High School, Columbia, Missouri 65203, USA}
\author{Matthew~Snyder}
\affiliation {Department of Physics and Astronomy, University of Missouri, Columbia, Missouri 65211, USA}
\author{Jacob~Cook}
\affiliation {Department of Physics and Astronomy, University of Missouri, Columbia, Missouri 65211, USA}
\author{Duy~Tung~Nguyen}
\affiliation {Department of Physics and Astronomy, University of Missouri, Columbia, Missouri 65211, USA}
\author{P.~V.~Sreenivasa~Reddy}
\affiliation {Department of Physics, National Cheng Kung University, Tainan 701, Taiwan}
\author{Tay-Rong~Chang}
\affiliation {Department of Physics, National Cheng Kung University, Tainan 701, Taiwan}
\author{Shengyuan~A.~Yang}
\affiliation {Research Laboratory for Quantum Materials, Singapore University of Technology and Design, Singapore, Singapore}
\author{Guang~Bian}
\affiliation {Department of Physics and Astronomy, University of Missouri, Columbia, Missouri 65211, USA}

\begin{abstract}

Two-dimensional (2D) Dirac states with linear band dispersion have attracted enormous interest since the discovery of graphene. However, to date, 2D Dirac semimetals are still very rare due to the fact that 2D Dirac states are generally fragile against perturbations such as spin-orbit couplings. Nonsymmorphic crystal symmetries can enforce the formation of Dirac nodes, providing a new route to establishing symmetry-protected Dirac states in 2D materials. Here we report the symmetry-protected Dirac states in nonsymmorphic $\alpha$-antimonene (Sb monolayer). The antimonene was synthesized by the method of molecular beam epitaxy. 2D Dirac states with large anisotropy were observed by angle-resolved photoemission spectroscopy. The Dirac states in $\alpha$-antimonene are spin-orbit coupled in contrast to the spinless Dirac states in graphene. The result extends the “graphene” physics into a new family of 2D materials where spin-orbit coupling is present.

\end{abstract}
\maketitle

Dirac states with linear band dispersion and vanishing effective mass have been discovered in condensed matter materials such as graphene, topological insulators, and bulk Dirac/Weyl semimetals. \cite{Novoselov2005, Kim2005, Kane2010, RevModPhys.90.015001}. The topological phase of Dirac fermion states leads to many exotic physical properties such as zero-energy Landau levels \cite{Castro2009}. The graphene-like Dirac materials have been considered as a cornerstone for the development of next-generation electronic devices. However, the two-dimensional Dirac cones are generally unstable because an energy gap can be easily induced by intrinsic or extrinsic perturbations such as spin-orbit coupling (SOC) and lattice distortions \cite{Zhou2008}. This leads to the rarity of 2D Dirac materials, and only a few materials have been proved experimentally to host gapless 2D Dirac states, including graphene \cite{Novoselov2005, Kim2005} and the surface of topological insulators \cite{Kane2010}. The Dirac states in graphene can be considered to be gapless only by ignoring the small SOC of the system. When the dimension of  topological insulators is reduced, a tunneling energy gap opens at the Dirac point, which is due to the hybridization of the surface states on the opposite surfaces \cite{Kane2010}. To find gapless Dirac states, geometrical constraints on the crystal lattice are generally needed to protect the nodal points against various gapping mechanisms. 

Recently, it has been proposed that nonsymmorphic crystalline symmetries including glide mirrors and screw axes can enforce band crossings and hence induce Dirac-fermion like states \cite{Young2015, Guan2017, Kane2016, Wieder2018, Po2017, Slager2013, PhysRevX.7.041069}. This is because the operators of nonsymmorphic symmetry operations create only high-dimensional irreducible representations at certain symmetry points of the Brillouin zone \cite{Kane2016, Po2017, Slager2013, PhysRevX.7.041069, Fang2015, Hourglass2016, Yang2018, PhysRevB.96.155206, PhysRevB.95.075135}. This presents a new route to realizing 2D Dirac materials whose Dirac states are robust even in the presence of strong SOC. Here we report the   observation of 2D Dirac states in the $\alpha$-phase of antimonene (monolayer antimony). Bulk Sb is a group-V semimetal with a small overlap between the valence and conduction bands. Despite small electron and hole pockets at the Fermi level, there exists an indirect negative energy gap traversing the whole Brillouin zone and separating the valence and conduction bands \cite{Liu1995}. This gapped band structure enables Sb, though semimetallic, to host the same ${\bf Z}_2$ topological invariants as topological insulators \cite{Kane2010, Bian2011}. The large spin-orbit coupling plays a key role in the generation of the nontrivial band topology. In the 2D limit, monolayer Sb, $i.e.$, antimonene, is known to have two allotropic structural phases, namely, the black-phosphorus (BP)-like $\alpha$-phase and the hexagonal $\beta$-phase. The lattice of $\alpha$-antimonene ($\alpha$-Sb for short) is nonsymmorphic, meaning that $\alpha$-Sb can host symmetry-protected Dirac states. In this work, we synthesized $\alpha$-Sb by the technique of molecular beam epitaxy (MBE) and detected the 2D Dirac states by angle-resolved photoemission (ARPES) experiments. The results shed light on the search of 2D Dirac materials in the vast territory of 2D nonsymmorphic crystals. 

 In our experiment, $\alpha$-Sb was grown on SnSe substrates under an ultrahigh vacuum environment. The SnSe crystals were cleaved $in~situ$ and provided an atomically flat surface for the deposition of Sb. The crystallographic structure of $\alpha$-Sb is shown in Figs~1(a) and 1(b). The surface unit cell is marked by a blue rectangular box. The in-plane lattice constants are 4.49~\AA~and 4.30~\AA~in the $x$ and $y$ directions, respectively. The in-plane nearest-neighbor bond length is 2.90~\AA. $\alpha$-Sb consists of two horizontal atomic sublayers.  Each atomic sublayer is perfectly flat according to the first-principles lattice relaxations. For a single layer (1L) of $\alpha$-Sb, the vertical spacing between the two atomic sublayers is 2.79~\AA. For a two-layer (2L) $\alpha$-Sb film, the vertical distance between the two atomic sublayers within each $\alpha$-Sb layer is 2.89~\AA~while the spacing between the two layers of $\alpha$-Sb is 3.18~\AA. The structure of one- and two-layer $\alpha$-Sb belongs to the $\#$42 layer group ($pman$). The lattice is nonsymmorphic, because it is invariant under a glide mirror reflection operation. The glide mirror is parallel to the $x$-$y$ plane and lies in the middle between the two atomic sublayers of 1L $\alpha$-Sb. The glide mirror reflection is composed of a mirror reflection and an in-plane translation by (0.5$a$, 0.5$b$), where $a = 4.30$~\AA~and $b = 4.49$~\AA~are the lattice constants in the $x$ and $y$ directions, respectively. For 2L $\alpha$-Sb, the glide mirror sits in the middle between the two $\alpha$-Sb layers. Figures~1(c-e) show the STM image of two $\alpha$-Sb/SnSe samples with atomic resolution. The first sample consists of mainly 1L $\alpha$-Sb islands, see Fig.~1(c). A line-mode reconstruction (Moir{\'e} pattern) can be seen on the $\alpha$-Sb surface, as marked by the green dashed lines. The height profile is taken along the blue arrow (shown in Fig.~1(d)) indicates that the height of the 1L $\alpha$-Sb island on the SnSe surface is 6.5~\AA. The second sample possesses both 1L and 2L domains as shown in Fig.~1(e).
 
 The glide mirror symmetry of the lattice leads to band degeneracy at high-symmetry points $\bar{\mathrm{X}}_1 = (\pi, 0)$ and $\bar{\mathrm{X}}_2=(0, \pi)$. A detailed analysis of the location of Dirac points can be found in the previous work  \cite{Kowalczyk2020}. We performed first-principles calculations for the band structure of 1L and 2L $\alpha$-Sb films. The ABINIT package \cite{Gonze2009, Gonze2005} and a plane-wave basis set were employed in the calculations. The energy cut is 400 eV. Relativistic pseudopotential functions constructed by Hartwigsen, Goedecker, and Hutter (HGH) were used \cite{PhysRevB.58.3641}. The SOC of the system is varied from 0 to 300\% by linearly scaling the relativistic parts of the Hamiltonian \cite{Bian_2013}. The calculated band structures of 1L and 2L $\alpha$-Sb are shown in Figs.~2(a) and 2(b). The Brillouin zone of $\alpha$-Sb is plotted in Fig.~2(c). Both 1L and 2L $\alpha$-Sb have a semiconducting behavior, which can be seen in the calculated density of states. There exist band crossings at points $\bar{\mathrm{X}}_1$ and $\bar{\mathrm{X}}_2$. The band degeneracy occurs for every band at these two high-symmetry points. Each band splits into two branches as it disperses away from $\bar{\mathrm{X}}_{1,2}$. Therefore, the band crossings at $\bar{\mathrm{X}}_{1,2}$ create 2D Dirac states. The Dirac points at $\bar{\mathrm{X}}_{1}$ and $\bar{\mathrm{X}}_{2}$ in the top valence band are marked by `D1' and `D2', respectively. The location of Dirac points is entirely determined by the underlying nonsymmorphic lattice symmetry. We note that the Dirac points in $\alpha$-Sb are away from the Fermi level. There are a multitude of ways to shift the Fermi level in 2D materials, such as electrostatic gating and chemical doping. Therefore, the Dirac states can be accessed in transport experiments when the Fermi level is appropriately tuned in this monolayer system.
 
 The Dirac bands at $\bar{\mathrm{X}}_{1,2}$ can be described by an effective $k\cdot p$ model constructed around each Dirac point. For D1 at $\bar{\mathrm{X}}_{1}$, the matrix representations of the symmetry operations are $T=-i\sigma_y\otimes \tau_0 K$ (time reversal), $\widetilde{M}_{z}=\sigma_{z}\otimes \tau_y$ (glide mirror reflection), $P=\sigma_0\otimes\tau_x$ (space inversion), and $M_x=-i\sigma_x\otimes\tau_x$ (mirror reflection with respect to a plane parallel to the $y$-$z$ plane), where $K$ is the complex conjugation operator, $\sigma_j$ and $\tau_j$ ($j=x,y,z$) are the Pauli matrices representing spin and orbital  degrees of freedom, respectively, $\sigma_0$ and $\tau_0$ are the $2\times 2$ identity matrices. These matrices of symmetry operators can be found in the standard reference \cite{Bradley}. Subjected to these symmetry constraints, the effective model in the vicinity of D1 expanded to linear order in the wave vector $k'$ takes the form of
\begin{equation}
\mathcal{H}(\bm k')=v_x k'_x(\cos\theta\ \sigma_x\otimes\tau_z+\sin\theta\ \sigma_0\otimes\tau_y)+v_yk'_y\sigma_y\otimes \tau_z,
\end{equation}
where the energy and the wave vector $\bm k'=(k'_x, k'_y)$ are measured from D1, the model parameters $v_x$ and $v_y$ are Fermi velocity in the $x$ and $y$ directions, respectively, and $\theta$ is a real parameter that depends on the microscopic details. The dispersion around D1 is $E=\pm \sqrt{v_{x}^{2} {k'}_{x}^{2}+v_{y}^{2} {k'}_{y}^{2}}$, which indeed corresponds to a linear spin-orbit coupled  Dirac cone. According to the first-principles bands, $v_x = 3.74 \times 10^5\ m/s$ and $v_y = 3.60 \times 10^5\ m/s$. The Dirac cone exhibits a minor anisotropy. The effective model for D2 at $\bar{\mathrm{X}}_2$ can be described in a similar way. With $T=-i\sigma_y\otimes \tau_0 K$, $\widetilde{M}_{z}=\sigma_z\otimes \tau_y$, $P=\sigma_0\otimes\tau_x$, and $M_x=-i\sigma_x\otimes\tau_0$, the effective Hamiltonian can be written as
\begin{equation}\label{Heff2}
\mathcal{H}(\bm k')=v_x k'_x\sigma_y\otimes \tau_z+v_yk'_y(\cos\theta\ \sigma_x\otimes\tau_z+\sin\theta\ \sigma_0\otimes\tau_y),
\end{equation}
where $v_x=2.16 \times 10^4\ m/s$ and $v_y=8.35 \times 10^5\ m/s$, according to the first-principles calculation.
A huge anisotropy is found in the Dirac bands at D2. To find the SOC effect on the Dirac states, we calculated the the band structure with various strength of SOC and extract the Fermi velocity at D1 and D2, see Figs.~2(d) and 2(e). Without SOC, the two bands of the Dirac cone become degenerate in the direction of $\bar{\mathrm{X}}_{1}$-$\bar{\mathrm{M}}$-$\bar{\mathrm{X}}_{2}$. This can be seen in Eqns.~(1) and (2): the terms depending on the spin matrices $\sigma_{x,y,z}$ vanish in the absence of SOC, leaving only one term with $\sigma_0$, which induces the band splitting in $\bar{\mathrm{X}}_{1}$-$\bar{\mathrm{\Gamma}}$ and $\bar{\mathrm{X}}_{2}$-$\bar{\mathrm{\Gamma}}$ directions. Consequently, the bands form a nodal line at the boundary of the Brillouin zone,
since the bands are degenerate along $\bar{\mathrm{X}}_1$-$\bar{\mathrm{M}}$-$\bar{\mathrm{X}}_2$, see Fig.~2{d}. In this sense, the nodal-line band structure in the absence of SOC also arises from the nonsymmorphic symmetry of the lattice. However, this band degeneracy is not robust against spin-orbit coupling. Turning on SOC, the nodal line is gapped everywhere except $\bar{\mathrm{X}}_1$ and $\bar{\mathrm{X}}_2$. The Fermi velocity $v_y$ at D1 and $v_x$ at D2 grows from zero as SOC increases, leading to the formation of Dirac cones. In other words, SOC transforms the system from a nodal-line system into a Dirac fermion state. It is worth noting that $v_x$ at D2 remains highly suppressed even at artificially enlarged SOC, leading to an anisotropic Dirac cone at D2. 
 
The ARPES result taken from the 1L $\alpha$-Sb sample is shown in Fig.~3. The photon energy is 21.2~eV. There are three prominent features on the Fermi surface, namely, one electron pocket at $\bar{\mathrm{\Gamma}}$ and two hole pockets near $\bar{\mathrm{X}}_{2}$, see Fig.~3(a). The central pocket disappears in the iso-energy contour at $E = -0.4$~eV, as shown in Fig.~3(b). The band spectrum taken along the $\bar{\mathrm{X}}_2$-$\bar{\Gamma}$-$\bar{\mathrm{X}}_2$ and $\bar{\mathrm{X}}_1$-$\bar{\Gamma}$-$\bar{\mathrm{X}}_1$ directions are plotted in Figs~3(c) and 3(d). The calculated band dispersion is overlaid on the ARPES spectrum for comparison. The second derivative of the spectra near $\bar{\mathrm{X}}_2$ and $\bar{\mathrm{X}}_1$ is shown in Figs.~3(e) and 3(f) for a better visualization of the band dispersion. The theoretical bands agree with the ARPES spectrum, especially both showing the band crossings at $\bar{\mathrm{X}}_1$ and $\bar{\mathrm{X}}_2$. We note that the MBE sample is slightly electron-doped due to the charge transfer between the film and the substrate, and the Fermi level of the calculated bands is shifted to march the ARPES spectrum. Figures~3(g) and 3(h) show the band dispersion around $\bar{\mathrm{X}}_1$ and $\bar{\mathrm{X}}_2$ in a perpendicular direction. In the $\bar{\mathrm{M}}$-$\bar{\mathrm{X}}_2$-$\bar{\mathrm{M}}$ direction, the two subbands dispersing away from the Dirac nodes are nearly degenerate,  which leads to a huge anisotropy in the Dirac band contours. The anisotropy of Dirac bands is less prominent in the $\bar{\mathrm{M}}$-$\bar{\mathrm{X}}_1$-$\bar{\mathrm{M}}$ direction. This is consistent with the calculated band structure. We note that the minor discrepancy between the ARPES spectrum and first-principles bands can be attributed to the substrate effects on the MBE samples. We note that the linear Dirac bands along the $\bar{\mathrm{X}}_2$-$\bar{\Gamma}$ are clearly resolved by ARPES, because the Dirac state at $\bar{\mathrm{X}}_2$ is isolated in energy from the other occupied states and thus largely unaffected by the presence of the other bands. By contrast, the Dirac point at $\bar{\mathrm{X}}_1$ is close to the other states in energy, and the Dirac bands are bent in the vicinity of the other bands. This explains the tangled spectrum taken along the $\bar{\mathrm{X}}_1$-$\bar{\Gamma}$ direction. The ARPES results taken from the 2L $\alpha$-Sb sample is shown in Fig.4. According to the STM characterization, the sample possesses 1L and 2L domains, therefore we can see contributions from 1L and 2L to the total ARPES spectrum. From a comparison with the calculated bands, we can identify the spectrum from 2L $\alpha$-Sb films. The ARPES results again show Dirac points (band degeneracy) at $\bar{\mathrm{X}}_1$ and $\bar{\mathrm{X}}_2$. The Dirac cone centered at $\bar{\mathrm{X}}_2$ exhibits a large anisotropy in the band dispersion. The ARPES results along with the first-principles band simulations unambiguously demonstrate the existence of 2D Dirac states in the nonsymmorphic $\alpha$-Sb films. 

In summary, Our ARPES measurements and first-principles calculations showed that $\alpha$-Sb hosts Dirac-fermion states at the high-symmetry momentum points $\bar{\mathrm{X}}_1$ and $\bar{\mathrm{X}}_2$. The band degeneracy at the Dirac points is protected by the nonsymmorphic symmetry of the lattice. The lattice symmetry guarantees that the Dirac states in $\alpha$-Sb are robust even in the presence of strong SOC. SOC actually plays an important role in the formation of Dirac states. Without SOC, the two branches of the Dirac cone become degenerate along $\bar{\mathrm{X}}_{1,2}-\bar{\mathrm{M}}$ directions and result in a nodal-line band structure \cite{Kowalczyk2020}. In this sense, the Dirac states of $\alpha$-Sb are spin-orbit coupled, in contrast to the spinless ones of graphene. Breaking the lattice symmetry, on the other hand, can lift the band degeneracy at Dirac points and yield gapped phases \cite{Kowalczyk2020}. This correspondence principle of lattice symmetry and nodal band structure applies to all 2D nonsymmorphic crystalline materials. The results open a door for exploring novel "graphene" physics in a rich material pool of nonsymmorphic materials with strong spin-orbit coupling.

\bibliography{Sb_PRB}

\newpage

\begin{figure}
\includegraphics[width=1\linewidth]{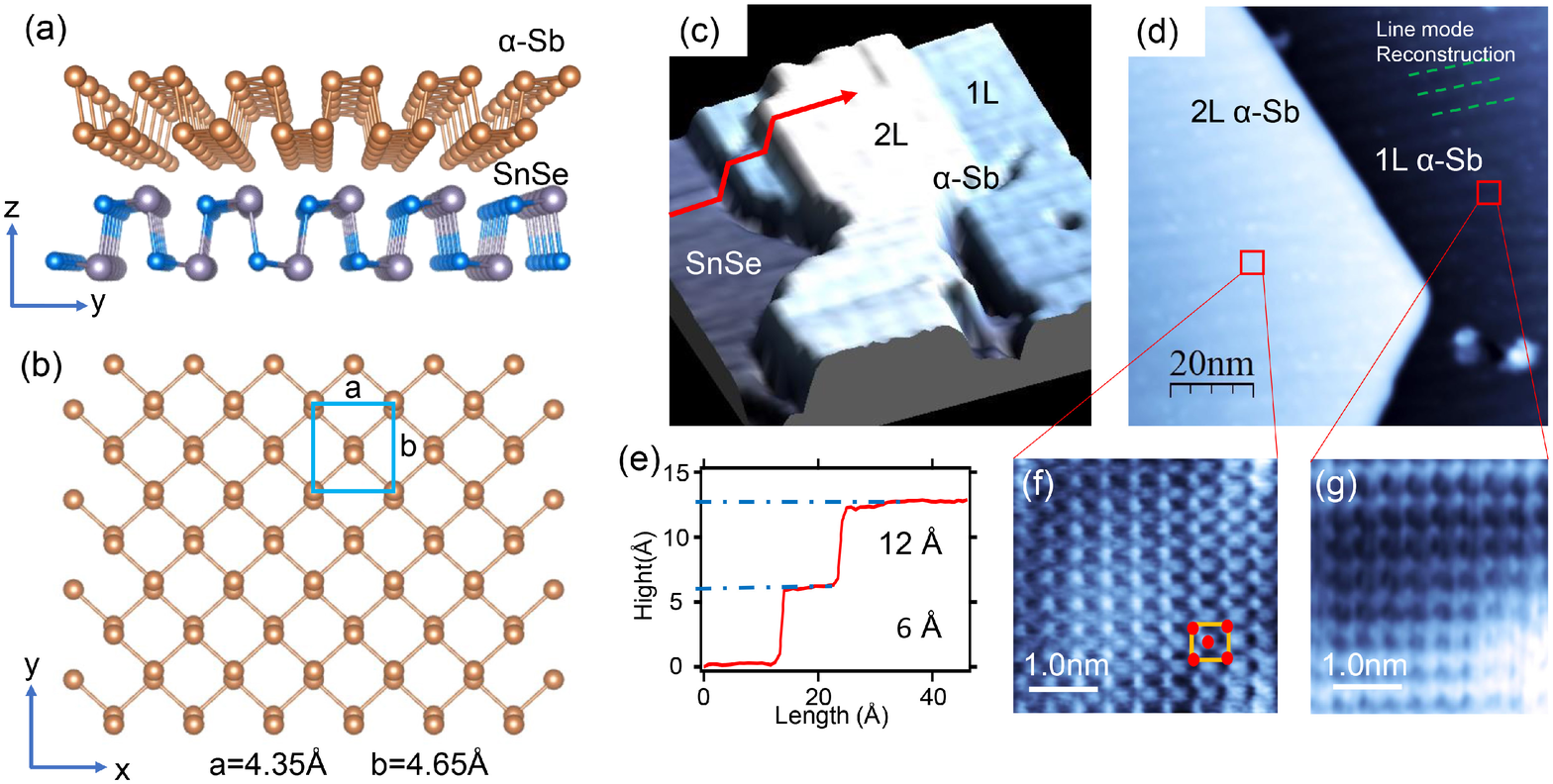}
\caption{(a) Side view of $\alpha$-Sb/SnSe lattice structure. (b) Top view of $\alpha$-Sb lattice structure. The unit cell is indicated by the blue rectangular box. The structure belongs to the $\#$42 layer group $pman$. (c) Surface morphology of an $\alpha$-Sb sample grown on SnSe substrate. (d) STM image of the $\alpha$-Sb sample showing 1L and 2L domains. (e) The height profile taken along the red arrow in (c). (f, g) Zoom-in STM images of the 1L and 2L domains shown in (d), respectively. 
}%
\end{figure}

\newpage

\begin{figure}
\includegraphics[width=1\linewidth]{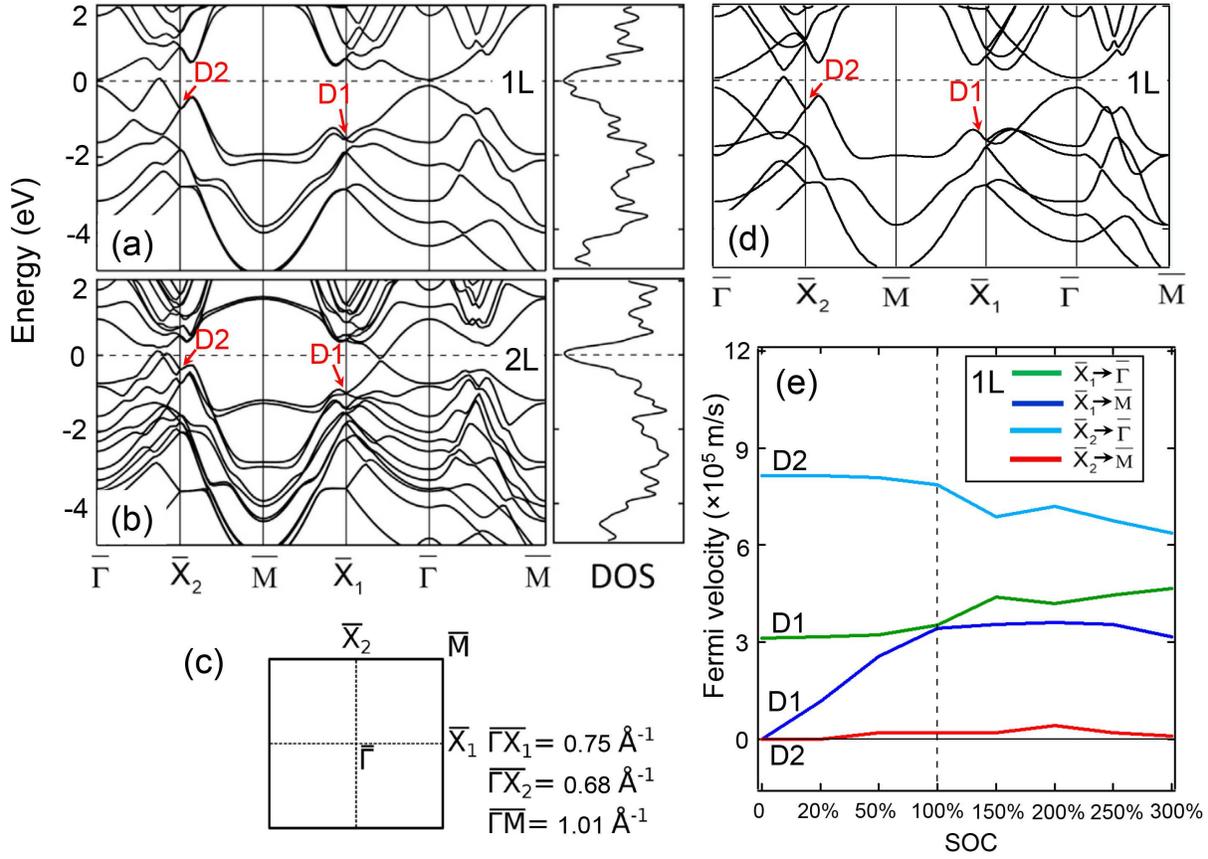}
\caption{(a) The band structure and density of states of 1L $\alpha$-Sb. (b) The band structure and density of states of 2L $\alpha$-Sb. (c) The Brillouin zone of $\alpha$-Sb. (d) The band structure of 1L $\alpha$-Sb without SOC. (e) Fermi velocity of the bands at D1 and D2 of 1L $\alpha$-Sb with various strength of SOC.}%
\end{figure}

\newpage
\begin{figure}
\includegraphics[width=1\linewidth]{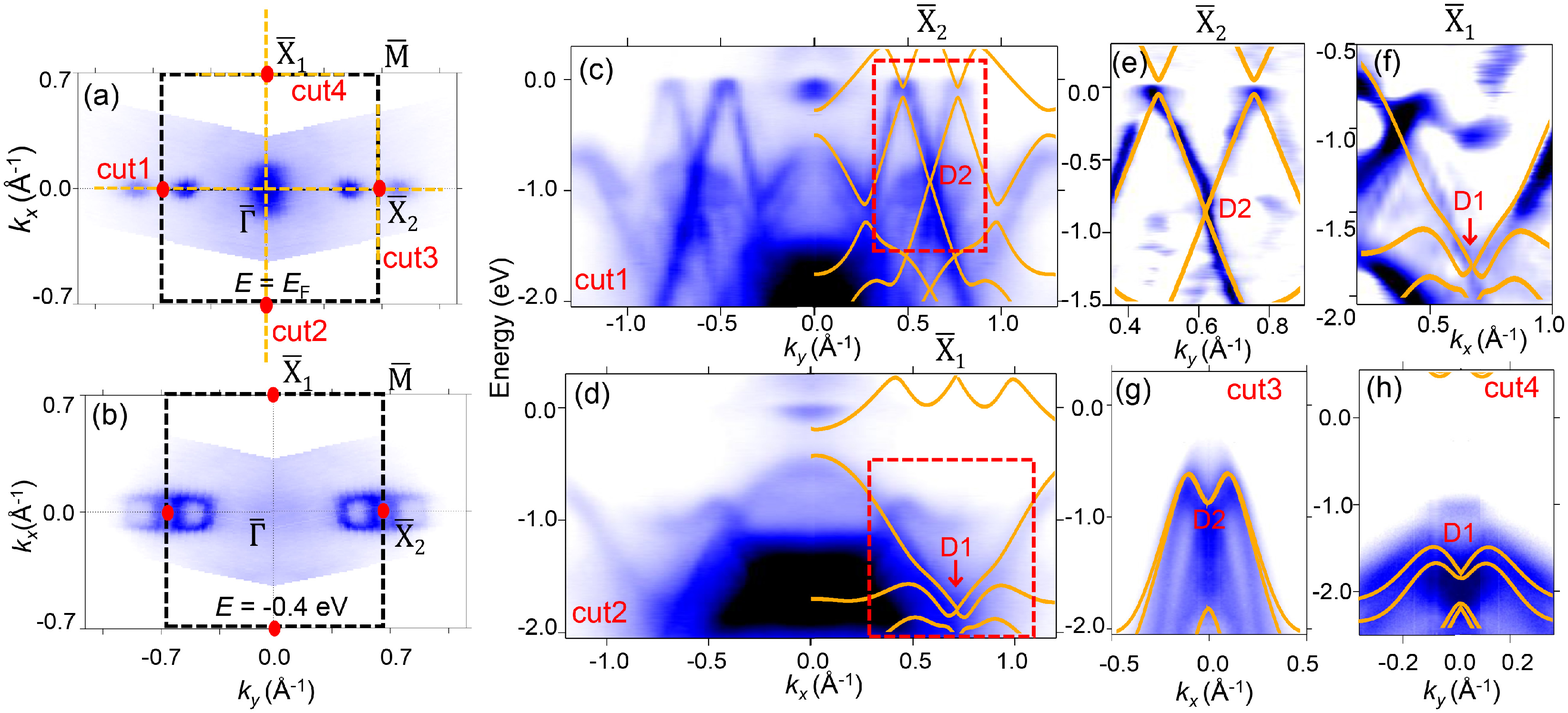}
\caption{ (a, b) Iso-energy contours of 1L $\alpha$-Sb taken at $E=E_{\rm F}$ and $-0.4~\rm{eV}$, respectively. (c, d) ARPES spectra taken along  the lines of `cut1' and `cut2' marked in (a), respectively.  The ARPES spectra are overlaid with the calculated bands. (e, f) Second derivative of the spectra shown in the red boxes marked in (c, d). (g, h) ARPES spectra taken along `cut3' and `cut4' marked in (a).}%
\end{figure}

\newpage
\begin{figure}
\includegraphics[width=1\linewidth]{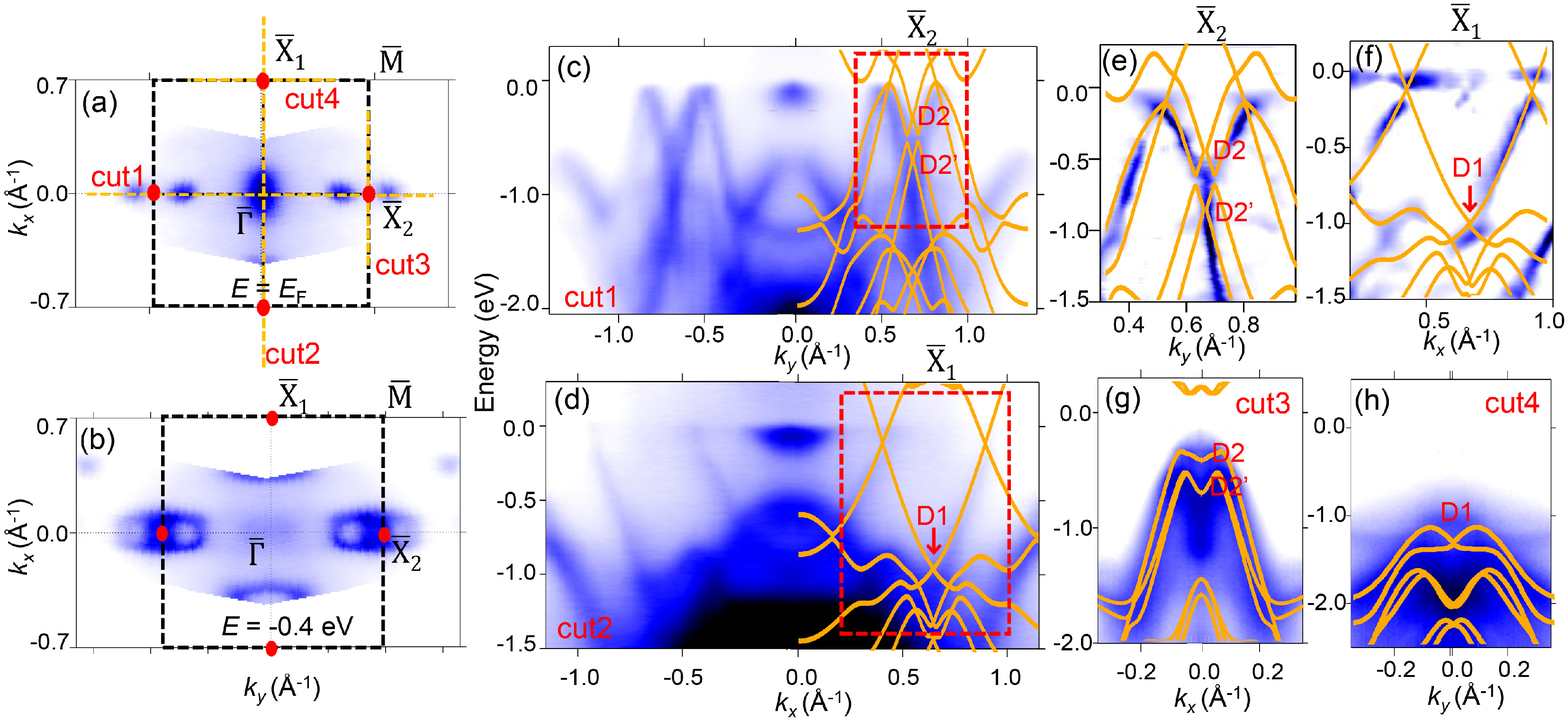}
\caption{ (a, b) Iso-energy contours of 2L $\alpha$-Sb taken at $E=E_{\rm F}$ and $-0.4~\rm{eV}$, respectively. (c, d) ARPES spectra taken along  the lines of `cut1' and `cut2' marked in (a), respectively.  The ARPES spectra are overlaid with the calculated bands. (e, f) Second derivative of the spectra shown in the red boxes marked in (c, d). (g, h) ARPES spectra taken along `cut3' and `cut4' marked in (a).}%
\end{figure}

\end{document}